\documentclass[oldversion]{aa}
\usepackage{times,graphicx,amssymb,lscape}
\usepackage{rotating}
\usepackage{natbib}
\bibpunct{(}{)}{;}{a}{}{,}  

\begin{document}

\title{Twisting Flux Tubes as a cause of Micro-Flaring Activity}

\author{D.B. Jess\inst{1,2}
        \and
        R.T.J. McAteer\inst{2}
	\and
        M. Mathioudakis\inst{1} 
        \and
        F.P. Keenan\inst{1}
	\and
	A. Andic\inst{1}
	\and
        D.S. Bloomfield\inst{3}
        }

\institute{ Astrophysics Research Centre, School of Mathematics and Physics, Queen's University, Belfast, BT7~1NN, 
Northern Ireland, U.K.
\and
NASA Goddard Space Flight Center, Solar Physics Laboratory, Code 612.1, Greenbelt, MD 20771, USA
\and
Max-Planck-Institut f\"{u}r Sonnensystemforschung, Max-Planck-Str. 2, 37191 Katlenburg-Lindau, Germany
}

\offprints{D.B. Jess, \email{djess01@qub.ac.uk}}

\date{Received 22 May 2007 / Accepted 13 September 2007}

\abstract{
High-cadence optical observations of an H-$\alpha$ blue-wing bright point near solar AR NOAA 10794 are 
presented. The data were obtained with the Dunn Solar Telescope at the National 
Solar Observatory/Sacramento Peak using a newly developed camera system, the {\sc{rapid dual imager}}. Wavelet analysis is undertaken 
to search for intensity-related oscillatory signatures, and 
periodicities ranging from 15 to 370~s are found with significance levels exceeding 95\%. During two separate microflaring events, 
oscillation sites surrounding the bright point are observed to twist. We relate the twisting of the oscillation sites to the twisting of 
physical flux tubes, thus giving rise to reconnection phenomena. We derive an average twist velocity of 8.1~km/s and detect a 
peak in the emitted flux between twist angles of $180^{\circ}$ and $230^{\circ}$.
}

\authorrunning{D.B. Jess et~al.}
\titlerunning{Twisting Flux Tubes as a cause of Micro-Flaring Activity}

\maketitle 
 
\section{Introduction}
\label{intro}
Magnetic reconnection remains one of the most promising mechanisms to convert magnetic energy into heating of the local plasma 
(see Klimchuk~2006). Indeed, recent theoretical work by De~Moortel~\& Galsgaard~\citep{DeM06a} suggests that photospheric footpoint 
motions lead to the build-up, and subsequent release, of magnetic energy. Two different types of footpoint motion have been proposed 
as potential mechanisms for reconnection, large-scale rotational motion and small-scale spinning. The proposed mechanism relies on 
two initially aligned, thin flux tubes, which are forced to interact due to footpoint motions. Twisting of the magnetic 
field in such a system will lead to the linear coupling of various MHD modes. However, the degree to which magnetic twisting modifies the 
dispersion relationship is not well understood (Nakariakov~\& Verwichte~2005).

\begin{figure}
\begin{center}
\includegraphics[angle=0,width=8.5cm]{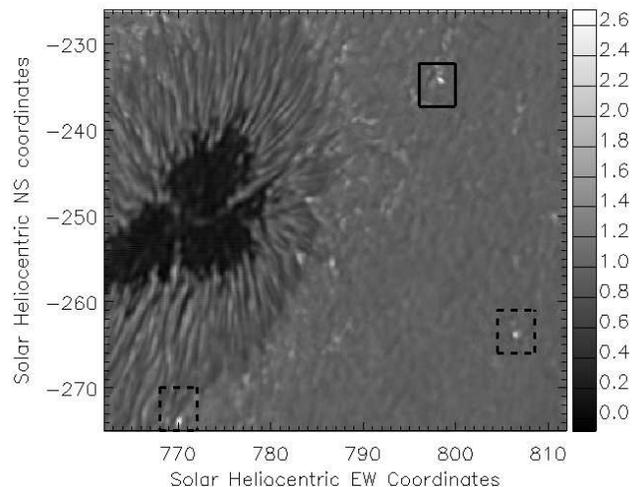}
\caption{\small The H-$\alpha$ blue wing field-of-view obtained at the DST in solar heliocentric coordinates. 
The colour scale indicates normalized flux and the solid box outlines the 
$40~\times~50$ pixel region removed for subsequent bright point analysis. The two dashed boxes outline additional bright points 
which were analysed for control purposes.}
\label{NBP}
\end{center}
\end{figure}

Recent work by Leenaarts~et~al.~\citep{Lee05} has 
discussed the formation of the H-$\alpha$ wing in detail using 3D magneto-convection simulations, and they show that H-$\alpha$ blue-wing 
bright points correspond to kiloGauss magnetic field concentrations in the photosphere. Co-aligned hard X-ray, H-$\alpha$ and magnetic 
field observations show that the X-ray emission of microflares occurs in small magnetic loops, while H-$\alpha$ emission originates 
from the footpoints of these loops located in the lower solar atmosphere (Liu~et~al.~2004). Due to the coincidence of these bright 
points with 
high magnetic field concentrations, magnetohydrodynamic (MHD) waves (Kalkofen~1997; Hasan \& Kalkofen~1999) appear to be the modes of 
wave propagation likely to be present (McAteer~et~al.~2002). Theory suggests that forms of MHD wave with longitudinal components to their
wave-vector can be induced via reconnection events (Hollweg~1981).

\begin{figure}
\begin{center}
\includegraphics[angle=90,width=8.5cm]{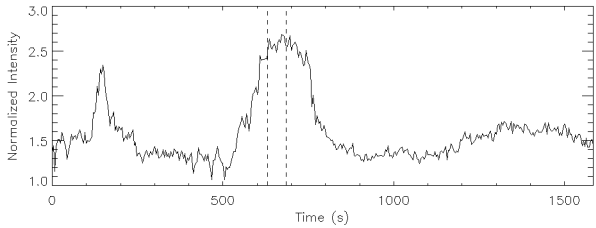}
\caption{\small The AR bright point lightcurve where the 
intensity is computed by averaging pixel values which supersede a lower-intensity threshold of the background modal value plus 
$10\sigma$. The vertical dashed line at 630~s corresponds to the time when the oscillatory power has undergone a $180^{\circ}$ twist. The 
second vertical dashed line at 685~s corresponds to the time when the oscillatory power has undergone a $230^{\circ}$ twist. Notice how 
the peak flux falls between the two vertical dashed lines in accordance with the theory in \S~\ref{results}.}
\label{lightcurve}
\end{center}
\end{figure}

If reconnection processes are to be both temporally and 
spatially resolved, it is imperative to acquire data at the highest possible cadence using a sensitive camera system.
Here we report intensity oscillations originating from an active region (AR) bright point, and the corresponding spatial rotation during 
microflare activity detected within the lower solar atmosphere.

\section{Observations}
\label{Obser}

The data presented here are part of an observing sequence obtained on 2005 August 10, with the Richard B. Dunn Solar Telescope (DST) 
at Sacramento Peak. Use of a highly sensitive, dual camera system enabled synchronized, simultaneous images at a rate of 
20 frames per second in each camera to be obtained. The {\sc{rapid dual imager}} ({\sc{rdi}}) camera system, used to acquire the 
observations presented here, was developed by Queen's 
University Belfast and remains stationed at the DST as a common user instrument. The {\sc{rdi}} system comprises of two 
Basler A301b cameras -- one `master' and one `slave' channel -- connected to a custom built PC, and controlled 
by software designed and developed by 4C's of Somerset, UK. The cameras have a 502 $\times$ 494 pixel$^2$ CCD, with a 
square pixel size of 9.9 $\mu$m and can operate at a maximum speed of 80 frames per second. Currently, the PC has 200~GB of 
available disk-space allowing for an uninterrupted data run of over 7 hours.

Our optical setup allowed us to image a $50.4\arcsec \times 49.2\arcsec$ region surrounding AR NOAA~10794 
complete with solar rotation tracking. The active region under investigation was located at heliocentric co-ordinates 
($770\arcsec$,$-254\arcsec$), or S12W56 in the solar NS-EW co-ordinate system. A Zeiss universal birefringent filter 
(UBF; Beckers~et~al.~1975) was used for H-$\alpha$ blue-wing (H-$\alpha$ core - 1.3\AA) imaging with one of the {\sc{rdi}} 
CCD detectors. The 6561.51\AA~H-$\alpha$ blue-wing central wavelength selected with the UBF had an associated filter bandpass of 
0.21\AA~(Beckers~et~al.~1975). In addition, a G-band filter was employed with the second {\sc{rdi}} camera to enable synchronized 
imaging in the two wavelengths. Due to the smaller granulation contrast of H-$\alpha$ blue-wing images when compared to G-band data,
only results obtained in the H-$\alpha$ blue wing are presented here. Leenaarts~et~al.~(2005) have demonstrated that, as a consequence 
of this reduction in granulation contrast for H-$\alpha$ blue-wing images, better feature tracking is obtained. During the observational 
sequence used in the analysis, low-order adaptive optics was implemented.

The data selected for the present analysis consist of 31760 H-$\alpha$ blue-wing images taken with a 0.05~s cadence over a total time 
period of 26.5~min. The acquisition time for this observing sequence was early in the morning and seeing levels were good with 
minimal variation throughout the time series. These images have a spatial sampling of $0.1\arcsec$ per pixel, to match the telescope's 
diffraction limited resolution in the H-$\alpha$ blue wing to that of the CCD. Further details regarding the {\sc{rdi}} camera system 
can be found in Jess~et~al.~\citep{Jes07}.

\begin{figure}
\begin{center}
\includegraphics[angle=0,width=8.5cm]{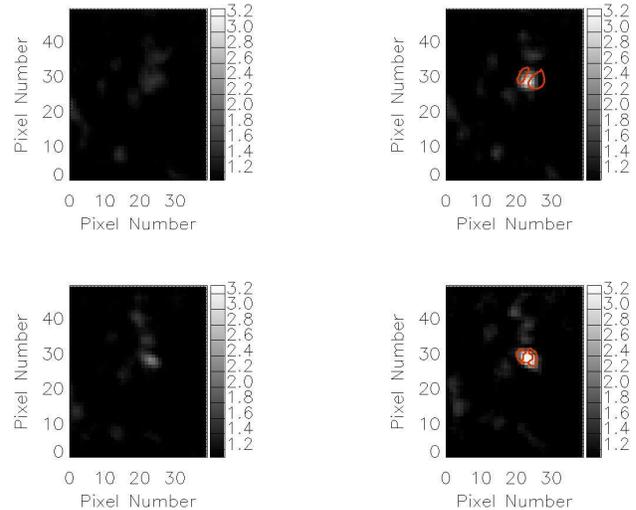}
\caption{\small The AR bright point before (upper left) and during (upper right) the first microflaring event. 
The lower~left panel displays the bright point prior to the second microflaring event, while the lower 
right panel shows the bright point undergoing the second microflaring event. Overplotted in each diagram (green contours) are locations of 
strong ($> 80$\% of the maximum of the entire time series) oscillatory power at a periodicity of 20~s. 
The microflaring bright point 
images reveal strong oscillatory power surrounding their centre as described in \S~\ref{results}.}
\label{superimposed}
\end{center}
\end{figure}

\section{Data Analysis}
\label{analy}

\begin{figure*}
\begin{center}
\includegraphics[angle=0,width=13.5cm]{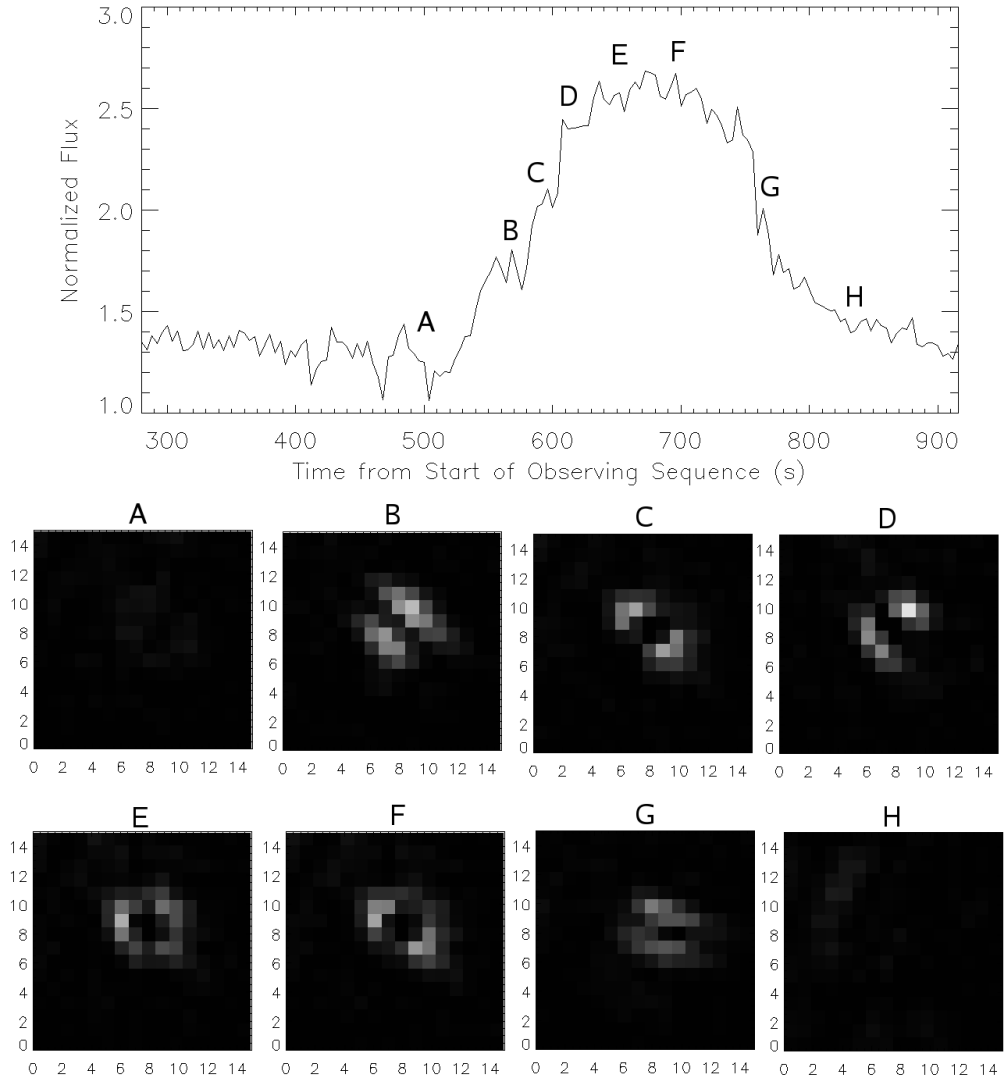}
\caption{\small The normalized flux of the bright point during the second microflare event is plotted, as a function of time, in the top 
panel. The bottom panel shows the spatial representation of oscillatory power for 20~s periodicities. The letters placed above each 
oscillatory power map indicate the time, through comparison with the top panel, at which the spatial wavelet-power image is generated. 
It can be seen that the twisting of the oscillation sites is closely linked to the brightening of the bright point. This figure is 
available as a movie for the online version.}
\label{twisting}
\end{center}
\end{figure*}

Small-scale turbulent seeing in the Earth's atmosphere means that even high-order adaptive optics cannot compensate for all 
rapid air motions.
Here we implement the speckle reconstruction masking method of Weigelt \& Wirnitzer \citep{Wei83}, adapted for solar imaging by von der 
L\"{u}he \citep{vdL93} and further improved by de Boer \citep{deB95}. Observing at a high cadence, the short 
exposure times essentially freeze out atmospheric distortions and maintain signals at high spatial frequencies, albeit with 
statistically disturbed phases (S\"{u}tterlin~et~al.~2001). It is possible to recover the true amplitudes and phases in Fourier 
space by taking a large number of such short exposure images, called a ``Speckle Burst'', and utilizing an elaborate statistical model.

\begin{figure}
\begin{center}
\includegraphics[angle=0,width=8.5cm]{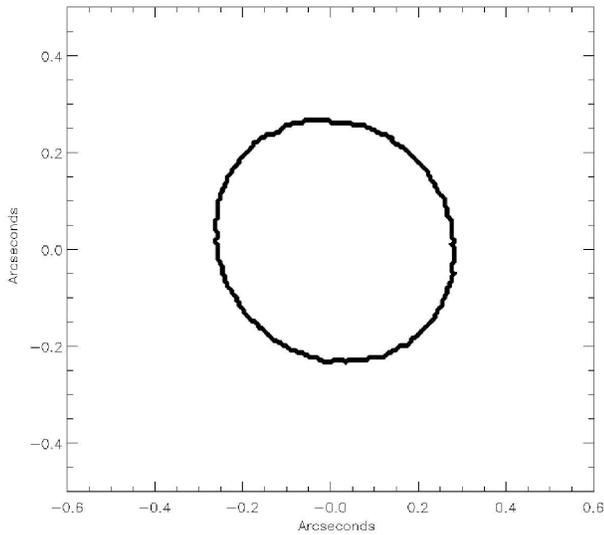}
\caption{\small A trace of the oscillation sites after de-rotation to solar disk centre. The path taken is not circular, providing a 
maximum traversed distance of 1230 km.}
\label{derot_path}
\end{center}
\end{figure}

Eighty raw {\sc{rdi}} data frames were used for each speckle reconstruction, producing a new effective cadence of 4~s. This provides a 
Nyquist frequency of 125~mHz and is therefore suitable for the search of oscillations with periods longer than 8~s. Typical Fried 
parameters obtained prior to speckle reconstruction were $r_{0}\approx$~10~cm, indicating good post-speckle image quality. 
Mikurda \& von der L\"{u}he~\cite{Mik06} have shown that although the residual wavefront errors differ between images obtained 
using low-order adaptive optics and uncompensated data, its effects on the speckle transfer function are, to a first-order 
approximation, similar to that of an uncompensated wave obtained during better seeing. Thus, since speckle reconstruction takes all 
residual aberrations into account, implementation of the speckle masking method outlined above is possible, on data obtained using 
low-order adaptive optics, without additional processing. However, 
after implementing temporal Fourier analysis on 2400 successive dark images, a moving-pattern noise was revealed, showing up as an intense 
Fourier power with a variable periodicity between 3 and 9~s. Therefore, to strengthen the reliability of our detections, only 
oscillations with periodicities greater than 15~s were incorporated into the analysis. Before commencing Fourier and wavelet techniques, 
a $40 \times 50$ pixel ($4\arcsec \times 5\arcsec$) region surrounding the AR bright point under consideration (Fig.~\ref{NBP}) 
was isolated for further 
study. All other spatial information was discarded to allow the analysis to focus on the behaviour of the bright point.  

To compensate for camera jitter and large-scale air-pocket motions, 
all data was subjected to a Fourier co-aligning routine. This routine utilizes 
cross-correlation techniques as well as squared mean absolute deviations to provide sub-pixel co-alignment accuracy. However, it 
must be noted that sub-pixel image shifting was not implemented due to the substantial 
interpolation errors which may accompany use of this technique. 
Since the observing sequence was obtained in the early hours of the morning, 
when image warping is particularly strong, all data were de-stretched relative to simultaneous, high-contrast G-band images. We use a 
$40 \times 40$ grid, equating to a $1.25\arcsec$ separation between spatial samples, to evaluate local offsets between successive 
G-band images. Due to both cameras sharing the same pre-filter optical path, all determined local offsets are applied to simultaneous 
narrowband images to compensate for spatial distortions caused by atmospheric turbulence and/or air bubbles crossing the entrance 
aperture of the telescope. However, even though wave front deformations in the pupil plane are first order wavelength independent, 
the image degradations are not, as can be seen when comparing derived Fried parameters at different wavelengths. Nevertheless, the 
de-stretching grid implemented in this process allows moderate- to large-scale seeing-induced distortions to be removed, since they are 
less wavelength dependent. With this in mind, very small-scale atmospheric distortions may not be accurately compensated for.

After successful co-alignment and de-stretching, time series were created for each pixel before being passed into Fast Fourier 
Transform (FFT) and wavelet analysis routines. 
The wavelet chosen for this study is known as a Morlet wavelet, and is the modulation of a sinusoid by a Gaussian envelope (Torrence~\& 
Compo~1998). A number of strict criteria implemented on the data allowed us to insure that oscillatory signatures correspond to real 
periodicities. These criteria have been described in detail in previous papers (see Banerjee~et~al.~2001, McAteer~et~al.~2004, 
Ireland~et~al.~1999, Mathioudakis~et~al.~2003).

Four-dimensional maps containing spatial information as well as wavelet power and oscillatory period were saved as outputs of 
wavelet analysis for detected oscillations which under the above criteria had a greater than 95\% significance level. Furthermore, a 
time series was created for the AR bright point to display its temporal variability throughout the time sequence. 
The bright point  
was identified by establishing which pixels in the $40 \times 50$ pixel region displayed intensity levels 
exceeding the background median value plus $10\sigma$. In each frame, approximately 30~pixels exceeded this threshold, 
indicating a bright point surface area greater than 150000~km$^2$. The 
pixels which superceded this intensity thresholding were averaged and placed at the corresponding time element in the time series array. 
Hence, variability in oscillatory phenomena can be compared readily with the bright point flux.

\begin{figure}
\begin{center}
\includegraphics[angle=0,width=8.5cm]{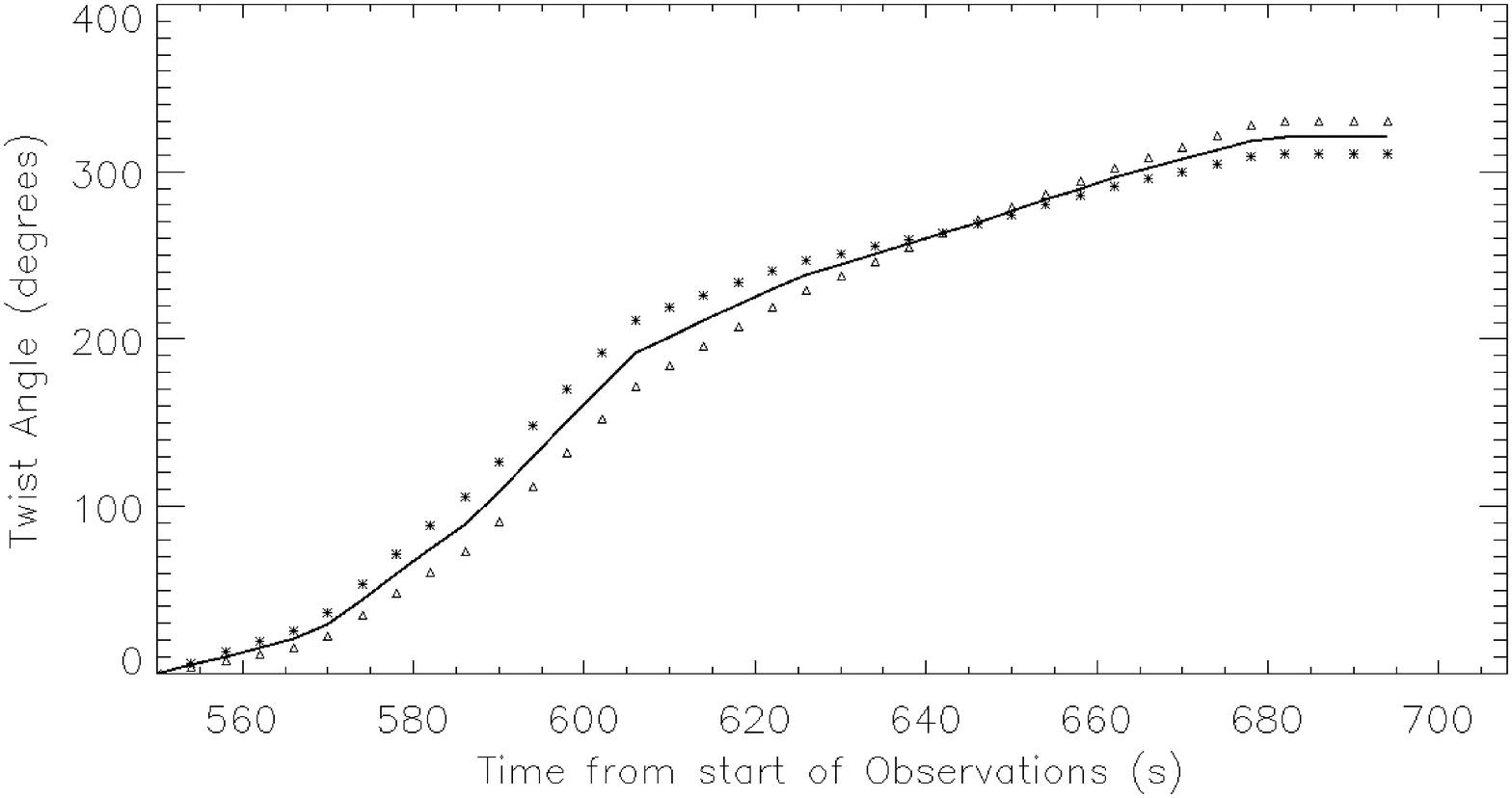}
\caption{\small The rotational angle of the footpoints as a function of time. The triangular and star symbols indicate the nature 
of the twisting for each of the oscillation sites, while the bold solid line shows the average rotational angle. As described 
in \S~\ref{results}, there is clearly an acceleration of the twist velocity between an angle of 20 and 160~degrees. Above 
160~degrees, the twist velocity decelerates continually until the motion ceases.}
\label{time_theta}
\end{center}
\end{figure}

\section{Results and Discussion}
\label{results}

An examination of the bright point time series over the complete 26.5~min duration of the observations reveals two events 
which show intensity increases  
(Fig.~\ref{lightcurve}). The first corresponds to an approximate 80\% intensity increase over the neighbouring quiescent flux, with an 
event duration exceeding 120~s, while the second is a much larger event and relates to an approximate 125\% increase in intensity 
lasting in excess of 350~s. Leenaarts~et~al.~\cite{Lee06} have shown that the brightness 
of H-$\alpha$ blue-wing features, just as for the G-band, will depend on the orientation of flux tubes, with higher intensities 
seen when the observer is looking straight down the tube axis. Thus, the swaying of flux tube bundles might explain temporary brightness 
enhancements when the buffeting of granules pushes the flux tubes, 
which are normally pointing 50 degrees away from the observer, towards them. However, due to the sharp rise in intensity and the 
corresponding decay pattern, we deem these intensity fluctuations to be characteristic 
of microflare activity with similar energetics and durations observed by Porter~et~al.~\citep{Por87}.

A succession of wavelet power diagrams around the time at which the microflaring activity occurs reveals a periodic 
signal. Prior to the commencement of both microflare events, there is an increase in oscillatory power superimposed, spatially, over 
the bright point (Fig.~\ref{superimposed}). The increase in oscillatory power provides two distinct regions of 
power separated by the central 
portion of the bright point. This increase in oscillatory power occurs 120~s prior to the second, larger microflare event which 
corresponds to 6~complete oscillation cycles for the 20~s periodicity plotted in Figure~\ref{superimposed}. During the evolution of both 
microflares, a spatial twist of the oscillatory sites occurs. The 
oscillation sites are not symmetric, yet appear to pivot around the bright point centre. Focusing on the second microflare event, 
Figure~\ref{twisting} reveals the nature of the twisting along with the associated time series.

\begin{figure}
\begin{center}
\includegraphics[angle=0,width=8.5cm]{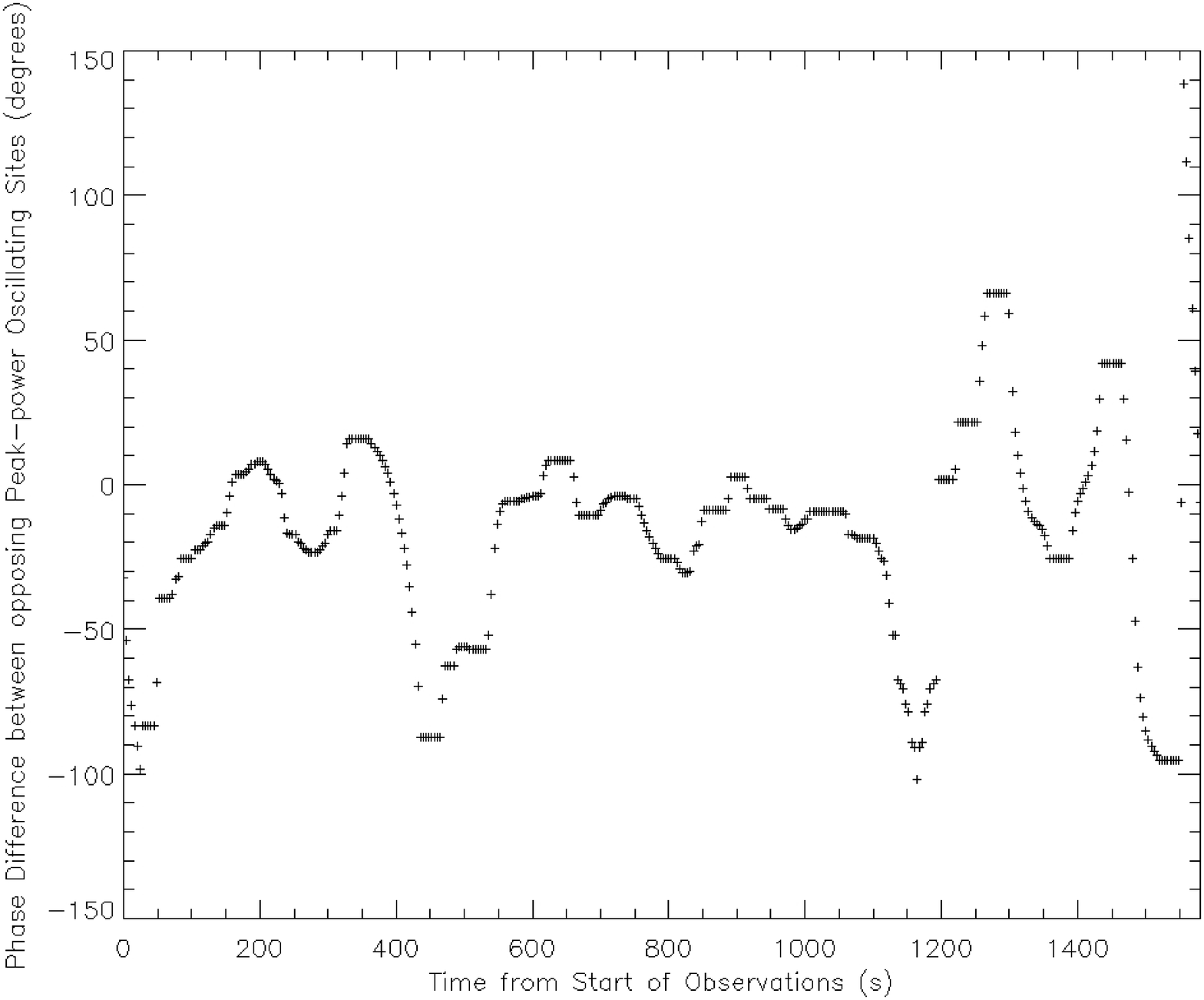}
\caption{\small Phase difference between the two oscillating sites, regardless of the magnitude of oscillatory 
power, prior, during and after the twisting is seen. Note the lack of anti-phase oscillations and minimal phase 
variation throughout the time series, particularly during the time interval when the second microflare event occurs (500-850~s).}
\label{phase}
\end{center}
\end{figure}

Due to the position of the active region near the West limb, line of sight effects must be considered. The observed circular path 
taken by the rotating oscillation sites will actually correspond to an elliptical path after 
consideration of the $cos\theta=0.62$ term. The path taken by the oscillating sites, after de-rotation to disk center, is shown in 
Figure~\ref{derot_path}. The perimeter distance around this path is 17~pixels, corresponding to a traversed distance of 1230~km, while 
the duration of twisting is 148~s. The rotational movement of the oscillatory twist fails to complete a full $360^{\circ}$ rotation, 
but we can use this distance to derive an upper limit to the average rotational velocity of the 
twisting oscillation sites, i.e. the twist velocity. Simply dividing the maximum traversed distance (1230~km) by the time taken (148~s), 
we establish a maximum average twist velocity of $\approx$ 8~km/s, very close 
to the photopsheric sound speed (Fossum~\&~Carlsson~2005). To demonstrate the temporal characteristics of the twist velocity, 
Figure~\ref{time_theta} plots the rotational 
twist angle as a function of time. From this figure it is clear that the rotational motion is accelerated between twist angles of 
$20^{\circ}$ and $160^{\circ}$, before being decelerated at rotational angles exceeding $160^{\circ}$. The twist angle is generated 
by tracking the wavelet power maxima as the oscillation sites pivot around the centre of the bright point. By establishing the 
movement of successive oscillatory power maximas, relative to the central pivot point, we are able to compute the twist angle 
directly. The high cadence of the observations, coupled with the high oscillatory power, provides a direct, and reliable, measure 
of the twist angle.

We have carried out a number of rigorous tests to assess the reliability 
of the observed twisting. The first test investigates the phase relationship of the two oscillating 
sites by comparing the positions of intensity peaks and troughs through use 
of a cross-power spectrum. Had the results indicated 
anti-phase behaviour, then the origin of the opposing oscillations would have been dubious. However, from Figure~\ref{phase}, it is clear 
that minimal anti-phase exists between the two oscillating sites, indicating 
real oscillatory phenomena. This 
finding is intuitive, since if the oscillation sites mark the footpoints of the same flux tube, any impulsive energy deposited, through 
reconnection, would produce oscillations which are coupled in phase (Nakariakov~et~al.~1999, Wang~et~al.~2003).

\begin{figure}
\begin{center}
\includegraphics[angle=0,width=8.5cm]{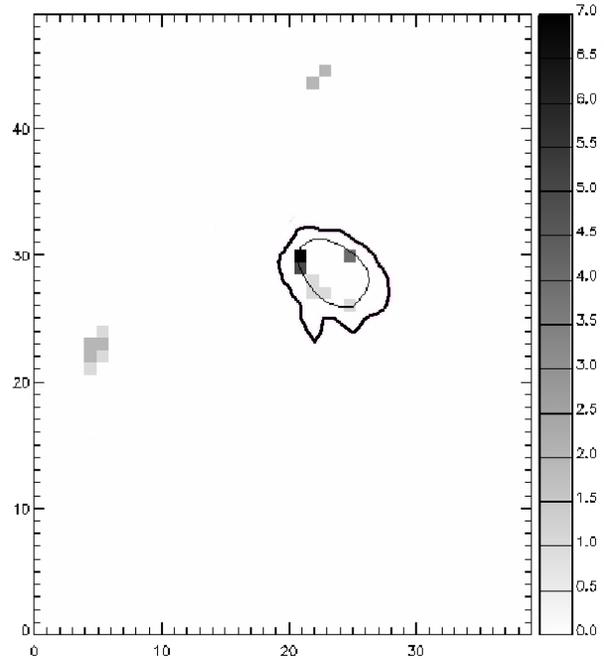}
\caption{\small The temporal evolution of the bright point (image scale in pixels). The colour scale indicates the number of times a 
pixel has disentangled itself from the confines of the bright point perimeter. The thin black 
contour outlines the average position and shape of the bright point during the entire observing sequence, while the thick black contour 
shows the time-averaged position of the oscillation sites. }
\label{outflows}
\end{center}
\end{figure}

\begin{figure*}
\begin{center}
\includegraphics[angle=0,width=15.2cm]{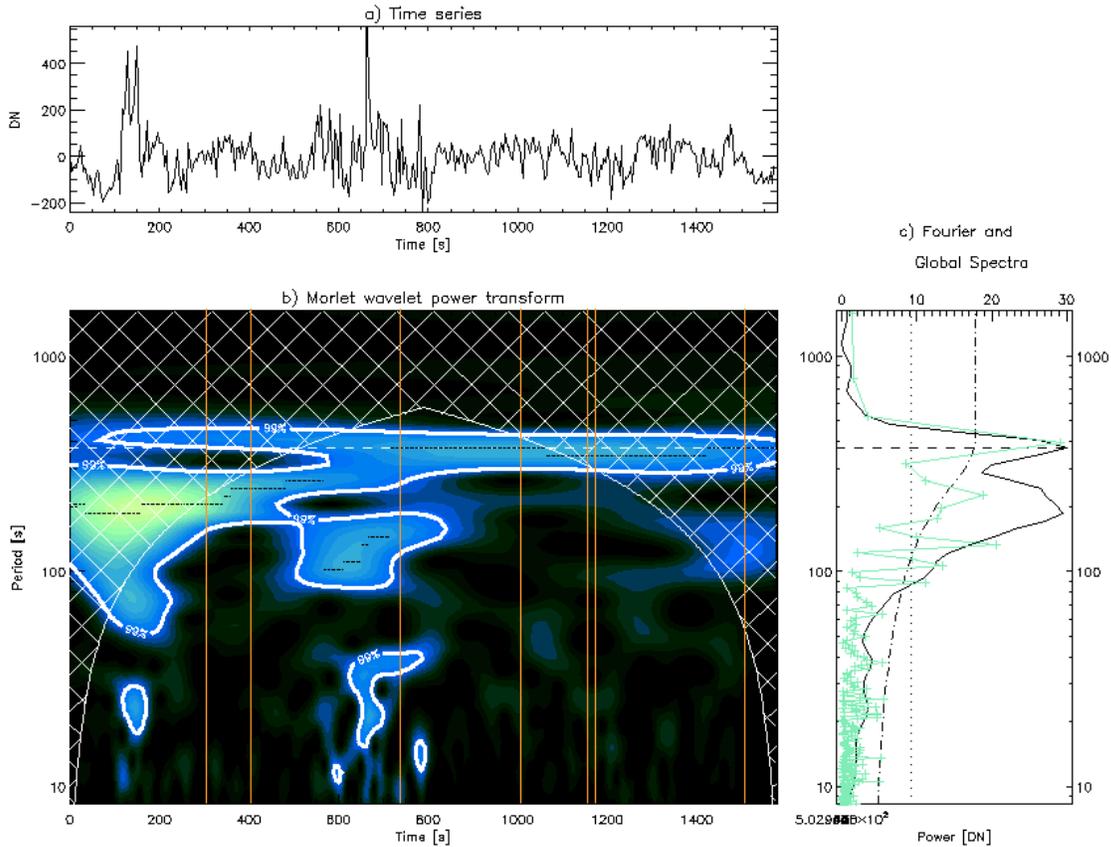}
\caption{\small Wavelet transform for the region of the bright point which undergoes the most temporal evolution during the 26.5~min 
time series. The original pixel lightcurve, normalized to zero, is shown in a), with the wavelet 
power transform along with locations where detected power is at, or above, the 99\% confidence level are contained within the contours 
in b). Plot c) shows the summation of the wavelet power transform over time (full line) and the Fast Fourier power spectrum (crosses) 
over time, plotted as a function of period. The global wavelet 
(dotted line) and Fourier (dashed dotted line) 95\% significance levels are also plotted. The cone of influence (COI), cross-hatched 
area in the plot, defines an area in the wavelet diagram where edge effects become important and as such any frequencies outside the 
COI are disregarded. Vertical lines indicate the times corresponding 
to outflows from the bright point. Notice how the outflows do not coincide, temporally, with the creation of high 20~s oscillatory power, 
which correspond to the micro-flare events described in \S~\ref{results}.}
\label{wavelet_time_overplot}
\end{center}
\end{figure*}

The second test creates a mask which maps the evolution of the bright point. Since intense pixels may move 
in and out of what we deem to be the perimeter of the bright point, this may produce large oscillatory power around its confines. We 
create a bright point pixel-mask for each frame of the time series, and by examining the image differences to 
previous masks, we are able to study the long-term evolution of the bright point. By summing all positive movements detected from 
successive masks, {\it{i.e.}} outflows, we can generate a picture of when, where, and by how much the bright point shape changes 
(Fig.~\ref{outflows}). 
After examination of Figure~\ref{outflows}, we can see some long-term evolution of the bright point, particularly around the 
North-Easterly edge where 7~outflows are detected over the entire 26.5~min duration of the time series. Overplotting the times related 
to these bright point outflows, on top of a wavelet power transform for the corresponding flux of the North-Easterly region, reveals no 
correlation with the locations of intense 20~s oscillatory power (Fig.~\ref{wavelet_time_overplot}). Furthermore, on closer inspection of 
Figure~\ref{outflows}, we see that the time-averaged position of the oscillation sites occupy a larger area than the bright point, 
indicating no dependence of the oscillation sites on outflows and/or jitter of the bright point.

\begin{figure}
\begin{center}
\includegraphics[angle=0,width=8.5cm]{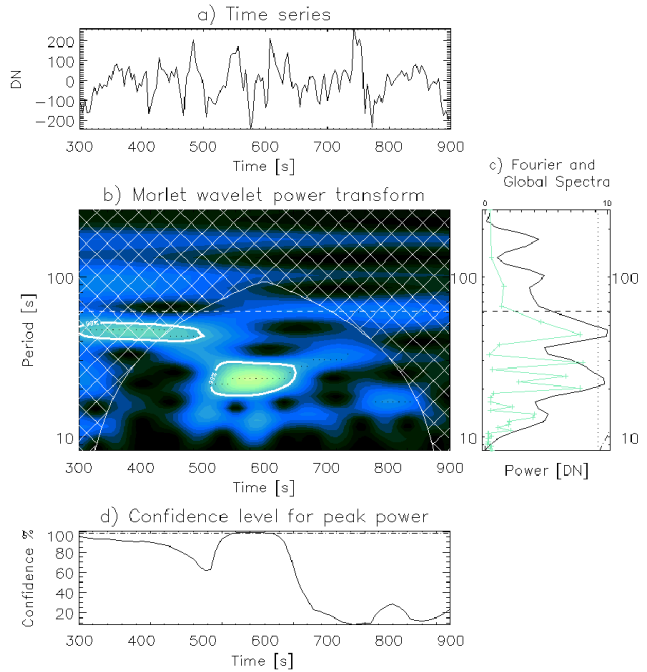}
\caption{\small Wavelet transform of the normalized flux of the bright point. All scales and axes are identical to 
Figure~\ref{wavelet_time_overplot}, except for the inclusion of a probability plot in d). The probability of detecting non-periodic 
power is calculated for the peak power at each timestep by comparing the value of power found in the input lightcurve with the number of 
times that the power transform of a randomized series produces a peak of equal or greater power. A percentage confidence is attributed 
to the peak power at every time step in the wavelet transform such that a low value of probability implies that there 
is no periodic signal in the data, while a high value suggests that the detected periodicity is real. We see that 
during the time of the second microflare event, strong power with high probability values, is associated with the 
detection of a 20~s periodicity.}
\label{brightpoint_20s_osc}
\end{center}
\end{figure}

Finally, our third test involves the examination of the bright point flux. Even though we see the oscillation sites pivot around 
the bright point centre, some of these oscillation sites overlap with its  confines. Thus, we should be able to detect 
intensity oscillations, albeit with much reduced power due to averaging over bright point pixels which are not oscillating, in the 
normalized flux output of the bright point. Running the bright point time series through wavelet analysis techniques, using the same 
strict criteria outlined in \S~\ref{analy}, we see that the 20~s oscillation is detectable (Fig.~\ref{brightpoint_20s_osc}). This enforces 
our belief that the detected oscillations are real, and we are not seeing an artifact of bright point evolution and/or jitter.

Leenaarts~et~al.~\cite{Lee06} have shown that the H-$\alpha$ wing is a great tool to locate and follow long-lived isolated magnetic 
elements, and since bright points in the blue wing of the H-$\alpha$ line correspond to kiloGauss magnetic 
field concentrations, magnetic features, such as flux tubes, may be present.
Introducing the theoretical work by De~Moortel~\& Galsgaard~\citep{DeM06a}, we can interpret the oscillatory sites described above as a 
bundle of initially aligned, thin flux tubes. Even though we cannot directly see the flux tubes, 
material flowing along such paths may induce magneto-acoustic oscillations, which we can detect with the wavelet analysis applied 
to the data. The De~Moortel~\& Galsgaard~\citep{DeM06b} model relates the evolution of the flux associated with magnetic reconnection to 
the rotational angle of initially aligned flux tubes. 

Interpreting our results as having a non-zero background potential field (consistent with same-polarity flux domains), the 
peak reconnection rate should occur between a 
twist angle of $180^{\circ}$ and $230^{\circ}$. Figure~\ref{lightcurve} shows the bright point time series during the second microflare 
event, along with vertical lines indicating the times where the twist angle equals $180^{\circ}$ and $230^{\circ}$. From this plot it is 
clear there is a peak in the emissive flux between these two angles. On closer examination of 
Figure~\ref{time_theta}, we can see that during peak acceleration, the twist velocity is greater than the sound speed. Between a time 
of 560 and 600~s, the oscillation sites are pivoted by approximately $160^{\circ}$, corresponding to a traversed distance of 540~km, and 
providing a velocity approaching 14~km/s. Kondrashov~et~al.~\cite{Kon99} have studied three-dimensional magnetohydrodynamic simulations 
of magnetic flux tubes, and have concluded that footpoints of interacting flux tubes, which are allowed to move to simulate conditions 
in the solar photosphere, promote the creation of supersonic motions during reconnection. We indeed see, from Figure~\ref{time_theta}, 
the development of supersonic motion after commencement of a microflare, which is consistent with the work by 
Kondrashov~et~al.~\cite{Kon99}.

Examining Figs.~\ref{time_theta}~and~\ref{lightcurve} more closely, 
it appears that the impulsive phase of the bright point microflare commences with a minimal twist angle. This is not in agreement with the 
model of De~Moortel~\& Galsgaard~\citep{DeM06b}, however, explanations can be proposed to help understand why this is the case. 
The model of De~Moortel~\& Galsgaard~\citep{DeM06b} relies on two initially aligned, line-tied plates between which the magnetic 
loops are not curved. This constrains the model with arbitrary parameters which may not accurately represent the observations presented 
here. Additionally, the distance of the plates between one another fully determines the 
degree of magnetic twist per unit length, and coupled with the loop curvature, predicts 
the starting time of reconnection. Furthermore, the arbitrary treatment of key input parameters such as diffusivity and pressure add to 
potential discrepancies between the model and the observations presented here. Thus, the model may not be advanced enough to accurately 
predict the minimal twist angle required for reconnection. Contrarily, twisting of the flux tubes 
may have commenced prior to our detection, with this discrepancy caused by our {\sc{rdi}} camera system having insufficient sensitivity 
to detect such weak signals. Current generation optical detectors should be able to address this problem with their improved dynamic 
ranges and sensitivities. This would allow weak rotational signals to be studied and more accurately compared with theory. Referring back 
to the first microflare event, a similar progression of oscillatory site evolution occurs. However, this event is less 
energetic, and as such our ability to trace exact movements is greatly reduced.

Two other bright points in the same field of view were also analysed using the same criteria described in \S~\ref{analy}. 
These bright points are of similar size and luminosity to the bright point in question (Fig.~\ref{NBP}). We find no evidence for flaring 
in these two locations and there is no twisting of oscillatory power.

\section{Concluding Remarks}
\label{conc}

We have presented direct evidence of high-frequency magneto-acoustic oscillations occurring in the immediate vicinity of an 
H-$\alpha$ blue-wing bright point. Periodicities as short as 15~s are found with significance levels greater than 95\%. We have 
interpreted the rotational movement of oscillatory sites during microflare 
activity as the physical signature of flux tube twisting. We derive the rotational angle corresponding to maximum flux and the average 
twist velocity. This observational result is in qualitative agreement with the numerical model introduced recently by 
De~Moortel~\& Galsgaard~(2006a,b). A direct quantitative comparison with the model is not possible due to uncertainties associated with 
key input parameters such as resistivity, diffusivity, pressure, the distance between opposing line-tied plates and the corresponding 
degree of magnetic twist per unit length.

\acknowledgements

This work was supported by the U.K. Particle Physics and Astronomy Research Council. DBJ wishes to thank the Department for 
Employment and Learning and NASA's Goddard Space Flight Center for a studentship. Doug 
Rabin and Roger Thomas deserve special thanks for their endless help, support and scientific input. FPK is grateful to AWE Aldermaston 
for the award of a William Penney Fellowship. Observations were obtained at the National Solar Observatory, operated by 
the Association of Universities for Research in Astronomy, Inc. (AURA), under cooperative agreement with the National Science Foundation. 
We would also like to thank an anonymous referee for very useful comments and suggestions. 
Finally we would like to thank the technical staff at the DST for perseverance in the face of atrocious weather. 
The construction of {\sc{rdi}} was funded by a Royal Society instrument grant. Wavelet 
software was provided by C. Torrence and G.P. Compo. 

~

\bibliography{aa}
\bibliography{submit}
 
\bibitem[2001]{Ban01}
Banerjee, D., O'Shea, E., Doyle, J. G., \& Goossens, M., 2001, A\&A, 371, 1137
\bibitem[1975]{Bec75}
Beckers, J. M., Dickson, L., \& Joyce, R. S., 1975, A Fully Tunable Lyot-\"{O}hman Filter (AFCRL-TR-75-0090; Bedford: AFCRL)
\bibitem[1995]{deB95}
de Boer, C. R., 1995, A\&AS, 114, 387
\bibitem[2006a]{DeM06a}
De Moortel, I., \& Galsgaard, K., 2006, A\&A, 451, 1101 
\bibitem[2006b]{DeM06b}
De Moortel, I., \& Galsgaard, K., 2006, A\&A, 459, 627
\bibitem[2005]{Fos05}
Fossum, A., \& Carlsson, M., 2005, Nature, 435, 919
\bibitem[1999]{Has99}
Hasan, S. S., Kalkofen, W., 1999, ApJ, 519, 899
\bibitem[1981]{Hol81}
Hollweg, J. V., 1981, Sol. Phys., 70, 25
\bibitem[1999]{Ire99}
Ireland, J., Walsh, R. W., Harrison, R. A., \& Priest, E. R., 1999, A\&A, 347, 355
\bibitem[2007]{Jes07}
Jess, D. B., Andi\'{c}, A., Mathioudakis, M., Bloomfield, D. S., \& Keenan, F. P., 2007, A\&A, accepted (arXiv:0707.2716v1 [astro-ph])
\bibitem[1997]{Kal97}
Kalkofen, W., 1997, ApJ, 486, 145
\bibitem[2006]{Kli06}
Klimchuk, J. A., 2006, Sol. Phys., 234, 41
\bibitem[1999]{Kon99}
Kondrashov, D., Feynman, J., Liewer, P. C., \& Ruzmaikin, A., 1999, ApJ, 519, 884
\bibitem[2005]{Lee05}
Leenaarts, J., S\"{u}tterlin, P., Rutten, R. J., Carlsson, M., \& Uitenbroek, H., 2005, ESA-SP, 596
\bibitem[2006]{Lee06}
Leenaarts, J., Rutten, R. J., S\"{u}tterlin, P., Carlsson, M., \& Uitenbroek, H., 2006, A\&A, 449, 1209
\bibitem[2004]{Cha04}
Liu, C., Qiu, J., Gary, D. E., Krucker, S., \& Wang, H., 2004, ApJ, 604, 442
\bibitem[2002]{McA02}
McAteer, R. T. J., Gallagher, P. T., Williams, D. R., Mathioudakis, M., Phillips K. J. H., \& Keenan, F. P., 2002, ApJ, 567, 165
\bibitem[2004]{McA04}
McAteer, R. T. J., Gallagher, P. T., Bloomfield, D.S., Williams, D. R., Mathioudakis, M., \& Keenan, F.P., 2004, ApJ, 602, 436
\bibitem[2003]{Mat03}
Mathioudakis, M., Seiradakis, J. H., Williams, D. R., Avgoloupis, S., Bloomfield, D. S., \& McAteer, R. T. J., 2003, A\&A, 403, 1101
\bibitem[2006]{Mik06}
Mikurda, K., \& von der L\"{u}he, O., 2006, Sol. Phys., 235, 31
\bibitem[1999]{Nak99}
Nakariakov, V. M., Ofman, L., DeLuca, E. E., Roberts, B., \&  Davila, J. M., 1999, Science, 285, 862 
\bibitem[2005]{Nak05}
Nakariakov, V. M., \& Verwichte, E., 2005, LRSP, 3
\bibitem[1987]{Por87}
Porter, J. G., Moore, R. L., Reichmann, E. J., Engvold, O., Harvey, K. L., ApJ, 323, 380
\bibitem[2001]{Sut01}
S\"{u}tterlin, P., Hammerschlag, R. H., Bettonvil, F. C. M., et~al., 2001, ASP Conference Series, Vol. 236
\bibitem[1998]{Tor98}
Torrence, C., \& Compo, G. P., 1998, Bull. Amer. Meteor. Soc., 79, 61
\bibitem[1993]{vdL93}
von der L\"{u}he, O., 1993, Sol. Phys., 268, 374
\bibitem[2003]{Wan03}
Wang, T. J., Solanki, S. K., Curdt, W., Innes, D. E., Dammasch, I. E., \& Kliem, B., 2003, A\&A, 406, 1105 
\bibitem[1983]{Wei83}
Weigelt, G., \& Wirnitzer, B., 1983, Optics Letters Vol. 8, No. 7, 389


\end{document}